\documentstyle{mn}
\newcommand{\be}{\begin{equation}}
\newcommand{\ee}{\end{equation}}
\newcommand{\ben}{\begin{eqnarray}}
\newcommand{\een}{\end{eqnarray}}

\title[apsidal corotation in 55 Cnc]{The apsidal corotation in mean motion resonance: \\the 55 Cancri as an
example}
\author[Zhou et al.]{Li-Yong Zhou$^{1,2}$\thanks{zhouly@nju.edu.cn},
        Harry J. Lehto$^{1,3}$, Yi-Sui Sun$^2$, Jia-Qing Zheng$^1$\\
    $^1$ Tuorla Observatory, V\"ais\"al\"antie 20, Piikki\"o 21500, Finland \\
    $^2$ Department of Astronomy, Nanjing University, Nanjing 210093, China\\
    $^3$ Department of Physics, Turku University, Turku 20500, Finland}

\begin{document}

\maketitle

\label{firstpage}

\begin{abstract}
The inner two planets around the 55 Cancri were found to be
trapped in the 3:1 mean motion resonance. In this paper, we study
the dynamics of this extra-solar planetary system. Our numerical
investigation confirms the existence of the 3:1 resonance and
implies a complex orbital motion. Different stable motion types,
with and without the apsidal corotation, are found. Due to the
high eccentricities in this system, we apply a semi-analytical
method based on a new expansion of the Hamiltonian of the planar
three-body problem in the discussion. We analyse the occurrence of
the apsidal corotation in this mean motion resonance and its
influence on the stability of the system.
\end{abstract}


\section{Introduction}
Over 100 extra-solar planets have been found (see e.g.
http://www.obspm.fr/encycl/encycl.html by J.Schneider). The high
eccentricities, close distances to the central stars and the heavy
masses, show the wide variety of planetary system that are quite
different from our solar system.

Among these ``exoplanets'', 28 are located around the main
sequence stars in 13 ``multiple planet systems'', and some of them
appear to be in Mean Motion Resonances (hereafter MMR). Recently,
the two inner planets around the 55 Cancri were found to be in a
possible 3:1 MMR \cite{jij03}. In numerical simulations, the
resonant angles are found to librate around certain values. At the
same time, the difference between the two periastrons is found to
be locked to a fixed value, that is, the two planets precess at
the same rate.

It is commonly accepted that these extra-solar planets are not
formed {\it in situ} but have suffered orbital migrations.
Different migration speeds of planets may lead to a variation of
mutual distance between two neighboring planets and eventually a
capture into the current resonant configurations
\cite{kle00,kle03,nel02}. The dynamics of these resonances may
contain important information of the capturing process and
therefore give valuable hints to the planet formation and the
evolution of a planetary disk. As a result, planetary systems with
resonance require special attention. Moreover, the possibility of
an Earth-type (habitable) planet in the 55 Cancri system has been
discussed \cite{cun03}. The existence of a habitable planet in a
system depends not only on the current dynamical features of the
system but also on the history it has experienced. Therefore it is
important to study the now-known planets before an Earth-like
planet was observed someday. All these points make this planetary
system interesting.

In this paper, we confirm the possibility that the two planets are
in a 3:1 MMR by numerical integrations and discuss the possible
configurations this system would take (section 2), then with a new
expansion of the Hamiltonian for a planar three-body problem
(section 3), we analyse the dynamics of this system. We will
discuss the occurrence of the apsidal corotation and its influence
on the stability of this system (section 4). Finally the
conclusions and discussions are given in section 5.

\section{Numerical simulations}
\subsection{Numerical model and initial set-up}
The 55 Cancri system is a system with three planets. We adopt the
orbital solution given by Fischer et al. \shortcite{fis03}, as
being listed in Table 1. To numerically simulate this system, we
use a symplectic integrator \cite{wis91,mik00}, which allows us to
follow the orbital evolution of each body and simultaneously the
stability of this orbit indicated by the Lyapunov Character
Indicator (LCI) \cite{fro84}. Each planet orbit has an LCI and we
choose the largest one among the three to indicate the stability
of this planetary system in this paper. The time step is set to be
$0.3$\,d, which is about $2\%$ of the orbital period of the
innermost planet.

\begin{table}
\caption{The orbital solution from Fischer et al (2003). The mass
of the central star is $1.03 M_{\sun}$, and in this paper we adopt
the planet masses by assuming $\sin i=1$.}
\begin{tabular}{lrrr}
\hline
Parameter &Companion B &Companion C &Companion D \\
\hline
$M\sin i$ $(M_J)$ & 0.83  & 0.20  & 3.69\\
$P$ (days) & 14.65 & 44.27 & 4780\\
$a$ (au) & 0.115 & 0.241 & 5.461\\
$e$    & 0.03  & 0.41  & 0.28\\
\hline
\end{tabular}
\end{table}

We adopt the masses, semimajor axes and eccentricities of planets
the values listed in Table 1. The initial orbital inclination of
companion D is set to be zero, while the companion B and C have
initial inclinations of $10^{-5}$ degrees. Other angles (the
ascending node, the periastron and the mean anomaly) are randomly
generated from $[0,2\pi)$. Four hundred simulations, with
different initial conditions, are integrated up to $10^6$ years.

Ignoring the companion D, we first integrate a system composed of
the central star and the companion B and C. Then, retaining the
initial conditions of these three bodies and adding the companion
D on an initial orbit with the given $a,e,i$ and randomly selected
angle elements, we integrate the four-body system. As the
simulations show, the companion D has nearly no influence on the
motion of the inner planets, although it is much heavier. This is
due to its large semimajor axis ($5.461$\,au). However, in order
to get reliable simulations for the real system, we report here
the numerical results from the four-body model.

During the integrations, if the distance between any two of the
planets became smaller than half of the criterion for the ``Hill
stability'' \cite{gla93}, the system was considered become
collapsed and the simulation was terminated.

Generally, the motion of a planet is dominated by the central star
and its orbit is a conic section with small deviations due to the
gravitational perturbation from other planet(s). This perturbation
can be described by the {\it disturbing function}, which can be
expanded in terms of the orbital elements. We use
$a,e,i,\varpi,\Omega,\lambda$ to denote the semi-major axis,
eccentricity, inclination, longitude of periastron, longitude of
ascending node, and mean longitude, respectively. The disturbing
function for a planet with mass $m_1$ and orbital elements
$(a_1,e_1,i_1,\varpi_1,\Omega_1,\lambda_1)$, perturbed by another
planet (indicated by subscripts `2'), can be written as
 \[ R=\mu_2\sum F(a_1,a_2,e_1,e_2,i_1,i_2)\cos\varphi. \]
Here $\mu_2={\cal G}m_2$ and $\varphi$ is a linear combination of
angles
 \[ \varphi=j_1\lambda_1+j_2\lambda_2+j_3\varpi_1+j_4\varpi_2+j_5\Omega_1+j_6\Omega_2 \]
where $j_1,j_2,\cdots,j_6$ are integers satisfying
$j_1+j_2+\cdots+j_6=0$.

Particularly, when the orbital periods satisfy $P_2/P_1\approx 3$,
that is, $\dot\lambda_1-3\dot\lambda_2\approx 0$ (dots denote the
time derivation), we have $\dot\varphi_{(j_1=1,j_2=-3)}\approx 0$
since generally $\dot\varpi_1, \dot\varpi_2, \dot\Omega_1,
\dot\Omega_2 \ll \dot\lambda_1, \dot\lambda_2$. Then in the
disturbing function $R$, those terms containing
$\lambda_1-3\lambda_2$ can not be eliminated by the averaging
technique and become the leading terms. And if the inclinations
$i_1,i_2$ are not too large, the leading terms in $R$ are of
$O(e_1^2), O(e_2^2), O(e_1e_2)$ and the corresponding angle
$\varphi$, now called {\it resonant arguments}, are
 $\theta_1=\lambda_1-3\lambda_2+2\varpi_1$,
 $\theta_2=\lambda_1-3\lambda_2+2\varpi_2$,
 $\theta_3=\lambda_1-3\lambda_2+\varpi_1+\varpi_2$.
Therefore the libration of any one of $\theta_{1,2,3}$ indicates
the two planets are in an ({\it eccentricity-type}) 3:1 MMR. We
note that no more than two of them are linear independent and
$\theta_3=(\theta_1+\theta_2)/2$ [see for example Murray \&
Dermott \shortcite{mur99} for more details].

Beside these resonant arguments, the relative apsidal longitude
$\Delta\varpi=\varpi_1-\varpi_2$ is also a critical argument often
to be discussed. We call the libration of $\Delta\varpi$ in an MMR
the {\it apsidal corotation}. In a 3:1 MMR, if both $\theta_1$ and
$\theta_2$ librate with small amplitudes, $\Delta\varpi=
\frac{1}{2}(\theta_1 - \theta_2)$ will also necessarily librate.
In this sense, the apsidal corotation is not independent and
should be different from the term of {\it apsidal resonance},
which is in the context of a secular perturbation.

Hereafter we use subscripts $1,2$ to label the orbital elements of
companion B and C respectively.

\subsection{Numerical results}
About one third ($133$ out of $400$) of the simulations collapse
during the integrations, and all the remainders survive for the
$10^6$\,yr integration. Most of the survivors have very short
($<10^2$\,yr) e-folding time $T_e$ ($=1/{\rm LCI}$, the reciprocal
of LCI), while 38 of them have $T_e \gid 10^3$\,yr, and they are
in this sense regarded as stable. In these stable systems, all the
three planets have final inclinations $\la 0\degr.1$, but those
unstable ($T_e< 10^3$\,yr) systems may have inclinations of
companion B and C as high as $\sim 20\degr$, that is, the stable
system prefers to retain to be coplanar. All the stable systems
are found to be associated with the 3:1 MMR between the two inner
planets.

\begin{figure*}
\vspace{8.0cm} \includegraphics{fig1.eps} \caption{The temporal evolution of
the critical angles $\theta_1, \theta_2, \theta_3$,
$\Delta\varpi$, and the eccentricities of planets $e_1, e_2$ (in
the bottom panel, the dashed curves indicate $e_2$). Case {\bf a},
{\bf b}, and {\bf c} are from different initial conditions. }
%\label{fig1}
\end{figure*}

\begin{figure}
\vspace{5.5cm} \includegraphics{fig2.eps} \caption{A typical temporal
evolution of the semimajor axes $a_1, a_2$ in a stable system. }
%\label{fig2}
\end{figure}

According to different configurations of $\theta_{1,2,3}$ and
$\Delta\varpi$, the stable systems can be divided into three
groups, with representative cases illustrated in Fig.\,1. In all
the cases, the companion D is in a steady motion (we don't
illustrate this here) and, the semi-major axes of companion B and
C vibrate with very small variations (an example is shown in
Fig.\,2).

Case {\bf a} (Fig.\,1) shows a strong resonance with $\theta_1$,
$\theta_2$ and $\theta_3$ librating around $215\degr$, $75\degr$
and $325\degr$, respectively. Simultaneously, the $\Delta\varpi$
also librates around $250\degr$. The well known symmetric
configurations of $\Delta\varpi$ librating around $0\degr$
(alignment) or $180\degr$ (anti-alignment) imply the conjunctions
of the two planets always occur at a certain position and this
attributes to the stability of the system \cite{lee02}. However
the asymmetric configuration with $\Delta\varpi$ librating around
angles other than $0\degr$ or $180\degr$, has been predicted by
Beaug\'e, Ferraz-Mello \& Michtchenko \shortcite{bea03a}.

Case {\bf b} differs from case {\bf a} in the behavior of
$\theta_1$, $\theta_3$ and $\Delta\varpi$, which circulate now.
The $\theta_2$ still librates around $80\degr$ with an amplitude
of $40\degr$. In fact, when $\theta_2$ librates with a small
amplitude, we may assume $\theta_2\approx const.$, that is,
$\dot\theta_2\approx 0$. From $\dot\theta_2= \dot\lambda_1 -
3\dot\lambda_2 + 2\dot\varpi_2 \approx 0$ we can derive
\[ \frac{n_2 -\dot\varpi_2}{n_1 -\dot\varpi_2} \approx \frac{1}{3}, \]
where $n_{1,2}=\dot\lambda_{1,2}$ are the mean motions of planets.
In a reference frame corotating with the periastron of the second
planet, the relative mean motions are $n_k^\prime = n_k -
\dot\varpi_2$ $(k=1,2)$, and from $n_2^\prime/n_1^\prime\approx
1/3$ we know that in this reference frame the conjunctions of
planets always happen at a certain direction. Meanwhile, given
$n_k\gg \dot\varpi_2$, we have $n_2/n_1 \approx
n_2^\prime/n_1^\prime$ and there is still a 3:1 relation between
the two orbital periods. In the 55 Cancri system, our numerical
results show $n_{1,2}$ are larger than $\dot\varpi_2$ by at least
two order of magnitudes.

Although the numerical simulations give more examples with
configuration as case {\bf a} than case {\bf b} (28 vs 7, see
Table 2), we notice that among the most stable sets with e-folding
time $T_e\sim 10^6$\,yr (the same order as the integration time),
there are two sets having configurations as case {\bf b} but only
one as case {\bf a}. This prevents us from concluding which type,
case {\bf a} or case {\bf b}, is more stable.

In case {\bf c}, $\theta_1,\theta_3$ and $\Delta\varpi$ circulate,
while $\theta_2$ librates around $180\degr$ with a quite large
amplitude of $160\degr$. As a result, it is less stable than the
above two cases. The only three examples with this configuration
in our simulations all have e-folding times $< 5\times 10^3$\,yr.

Finally we summarize the e-folding times and the motion
configurations of the 38 stable systems in Table 2.

\begin{table}
\caption{The distribution of $T_e$ of the stable systems. The
numbers with subscripts {\bf a}, {\bf b} or {\bf c} are the sums
of systems with different configurations as shown in Fig.\,1.}
\begin{tabular}{lcccc}
\hline
$T_e$ (yr) & $\sim 10^3$ & $\sim 10^4$ & $\sim 10^5$ & $\sim 10^6$ \\
\hline
 Number & $12_{\bf a}+2_{\bf b}+3_{\bf c}$ & $8_{\bf a}+1_{\bf b}$ & $7_{\bf a}+2_{\bf b}$ & $1_{\bf a}+2_{\bf b}$ \\
 \hline
\end{tabular}
\end{table}

Besides the angles $\theta_{1,2,3}$ and $\varpi$, we notice that
the orbital eccentricities $e_1, e_2$ also have remarkably
different behaviors in these three cases. Another noteworthy
feature of numerical results is that the system has a symmetry
over angles $\theta_i$ and $\Delta\varpi$, that is, when $\phi$ is
a libration center of $\theta_i$ or $\Delta\varpi$, the argument
$(360\degr - \phi)$ is too. For example, we observe $\theta_2$
librates around $\sim80\degr$ and $\sim280\degr$ with nearly the
same ratio in our simulations.

\subsection{Some extended calculations}

To integrate all the surviving systems to $10^7$\,yr or even
longer is out of our current computational capability. However, to
explore further the stabilities of different configurations, we
integrate the three examples shown in Fig.\,1 further to $3\times
10^7$ years. Both the case {\bf a} and {\bf b} last this time span
and keep the configurations all the time. But case {\bf c} loses
the stability when the inclinations of companion B and C increase
to $\sim 5\degr$ after $\sim 1 \times 10^7$ years. As in those
unstable examples, this inclination increasing is due to a
sequence of close encounters between the two planets. On the
contrary, the mechanism of the 3:1 MMR can protect those systems
as case {\bf a} and {\bf b} from catastrophic close encounters.

Other 5 examples randomly selected from the 17 systems with
$T_e\sim 10^3$\,yr are integrated to $1\times 10^7$\,yr. Again,
except the one having motion configuration as case {\bf c}, they
all survive and have new e-folding time $T_e>10^3$\,yr. [The LCI
will converge to a certain value as the integrating time
approaches to infinity, so $T_e$ of an orbit would change when
it's integrated to a longer time span. see for example
\cite{fro84}.] As for those systems with larger $T_e$, they are
more stable and should have longer surviving lives.

So far, our discussions are all based on the small initial
inclinations of the two inner planets. We find that in stable
systems the inner two planets are locked to the 3:1 MMR and their
small orbital inclinations can be kept to at least $10^6$\,yr.
While in those unstable systems, chaotic close encounters of
planets could cause inclinations to increase. Furthermore, we also
numerically simulated hundreds of examples with larger initial
inclinations. In these systems, except the inclinations of the two
inner planets are set to higher values, all the orbital elements
are initialized in the same way as before, and we integrate these
systems to $10^6$\,yr. A brief summary is given in Table 3.

\begin{table}
\caption{The percentages of the survival and stable systems in the
simulations with different initial inclinations ($i^0_1= i^0_2$).}
\begin{tabular}{lccccccc}
\hline
$i^0_{1,2}$ (deg.) & $10^{-5}$ & 2.5 & 5 & 10 & 15 & 20 & 25 \\
\hline
 $P_{\rm survival}$ & 66 & 62 & 52 & 35 & 23 & 23 & 21  \\
 $P_{\rm stable}$ & 10 & 10 & 10 & 10 & 6 & 5 & 6 \\
\hline
\end{tabular}
\end{table}

We see the survive probability drops from $66\%$ to 21\% as the
initial inclinations increase from $10^{-5}$ to 25 degrees, but at
the same time the percentage of stable systems even does not
decrease until $i^0_{1,2}=10\degr$ before it attains a value of
$\sim 5\%$. This implies that the stability of the system is not
so sensitive to the inclinations, and if smaller than 10 degrees,
they have nearly no influence on the stability of the system. On
the other hand, that all the stable examples are associated with
the 3:1 MMR reveals again the importance of this resonance in
stabilizing this planetary system, even when the orbits are far
from coplanar. We don't give more analysis to the problem of
higher inclinations here, but it deserves a detailed discussion in
future .

\section{Analytical model}
To understand the dynamics of the 3:1 MMR in the 55 Cancri, it is
useful to describe it with an analytical model. As we mentioned
above, the companion D has a very weak influence on the motion of
the inner planets. In addition, the numerical simulations show it
is mainly the 3:1 MMR that is responsible for the stability of the
system, in spite of the higher orbital inclinations. So, a planar
three-body problem is a reasonable model for describing this
planetary system, especially when we mainly focus on the 3:1 MMR.
Of course, considering the possible substantial difference between
the coplanar and non-coplanar motion, a more realistic and
reliable model is still needed. On the other hand, the usual
analytical perturbation models, such as the Laplace expansions,
are not convergent for high eccentricities, and expansions with
low-order truncations may yield imprecise results \cite{fer94}.
Recently, a new expansion of the Hamiltonian suitable for
high-eccentricities has been developed by Beaug\'e and Michtchenko
\shortcite{bea03b}. We will apply this expansion to the system the
55 Cancri.

\subsection{The Hamiltonian}
Suppose three bodies with mass $M_0, m_1, m_2$ orbiting around
their common center of mass. $M_0\gg m_{1,2}$ is the central star
and $m_{1,2}>0$ are the planets. Near the $3:1$ MMR, a set of
canonical variables are defined as:
 \ben
 \lambda_1; \hspace{3.7cm}
 J_1=L_1+\frac{1}{2} (I_1+I_2)\nonumber \\
 \lambda_2; \hspace{3.7cm}
 J_2=L_2-\frac{3}{2} (I_1+I_2) \nonumber\\
 \sigma_1=\frac{3}{2}\lambda_2-\frac{1}{2}\lambda_1-\varpi_1; \hspace{0.9cm}
 I_1=L_1-G_1 \hspace{1.1cm} \\
 \sigma_2= \frac{3}{2}\lambda_2-\frac{1}{2}\lambda_1-\varpi_2; \hspace{0.9cm}
 I_2=L_2-G_2 \nonumber \hspace{1.1cm}
 \een
where $L_i=m_i^\prime\sqrt{\mu_ia_i}$ and $G_i=L_i\sqrt{1-e_i^2}$
are Delaunay variables, $a_i, e_i, \lambda_i, \varpi_i$ are the
orbital elements of the $i$th planet, $\mu_i = {\cal G}(m_i+M_0)$,
${\cal G}$ is the gravitational constant, and $m_i^\prime =
\frac{m_iM_0}{m_i+M_0}$ is the reduced mass. These canonical
variables define a four degree-of-freedom system. And let's bear
in mind $\theta_i=-2\sigma_i$ ($i=1,2$),
$\theta_3=-\sigma_1-\sigma_2$ and $\Delta\varpi = \sigma_2
-\sigma_1$.

After considering the conservation of angular momentum and
averaging over the fast-angle, we get two integrals of motion
$J_{\rm sum}= J_1+ J_2$, $J_{\rm dif}= J_1-J_2$. At the same time,
the problem is reduced to a two degree-of-freedom system in
canonical variables ($I_i,\sigma_i$). The resulting Hamiltonian
reads
 \be
 H=-\sum_{i=1}^2 \frac{\mu_i^2 m_i^{\prime 3}}{2L_i^2}+H_1
 \ee
 where $H_1$ is the truncated disturbing function,
 \ben
 H_1= \frac{{\cal G}m_1m_2}{a_2}\sum_{n=0}^3 \sum_{j=0}^{j_{\rm max}}
 \sum_{k=0}^{k_{\rm max}} \sum_{u=0}^{u_{\rm max}} \sum_{l=-l_{\rm
 max}}^{l_{\rm max}}R_{n,j,k,u,l}(\alpha-\alpha_0)^n \nonumber \hspace{-1.5cm} \\
 \times e_1^j e_2^k \cos(2u\sigma_1+l(\sigma_2-\sigma_1)).
 \een
Here $\alpha= a_1/a_2$ is the ratio between the two semimajor axes
and $\alpha_0$ is its value in the exact MMR. The coefficients
$R_{n,j,k,u,l}$ can be determined beforehand \cite{bea03b}.
Considering the high value of eccentricities, we adopt the upper
limits of sums $u_{\rm max}=l_{\rm max}=12$ and $j_{\rm
max}=k_{\rm max}=15$ in this paper. For convenience, the
Hamiltonian has been expressed explicitly in $(a_i, e_i)$ instead
of $(I_i, J_i)$, therefore the integrals $J_{\rm sum}$ and $J_{\rm
dif}$ don't appear explicitly but exist as constrains. We note the
total energy $H$ is an integral too. Furthermore, it is easy to
prove the symmetry buried in the Hamiltonian:
 \[ H(\sigma_1,\sigma_2) \equiv H(-\sigma_1, -\sigma_2) \equiv
 H(180\degr+\sigma_1, 180\degr +\sigma_2). \]

\subsection{The reliability of the Hamiltonian}
We can calculate planet orbits by applying this Hamiltonian. Since
it is an averaged system, the time step can now be much longer
than that used in our symplectic integrator. A comparison between
this Hamiltonian expansion and the direct numerical integration is
shown in Fig.\,3, in which we choose the most unstable case of
Fig.\,1c as an example. We get the same initial conditions and
recalculate the orbits from the Hamiltonian equations with an
average (variable) time step of $60$\,d, which is four times of
the inner planet's orbital period.

\begin{figure}
\vspace{10.0cm} \includegraphics{fig3.eps} \caption{The temporal evolution of
$\theta_2$, $\Delta\varpi$ and eccentricities $e_{1,2}$ from the
same initial conditions as the ones of Fig.\,1c. Squares show the
results of a numerical integration of the exact dynamical
equations for a 4-body model, while the curves are from the
Hamiltonian equations for a 3-body model.}
%\label{fig3}
\end{figure}

The agreements in Fig.\,3, not only imply the reliability and
suitability of this Hamiltonian, but also reflect that the third
planet can be ignored when studying the dynamics of the inner two
planets.

\section{The apsidal corotation and the stability of the system}
The Hamiltonian introduced above has two degrees of freedom.
Although it is still non-integrable, it can help us to understand
the dynamical properties of the system. In this section, we will
use it to find out why and when the system follows the different
types of resonant evolutions as shown in Fig.\,1, and then via the
surface of section technique we study the stability of the systems
with different configurations.

\subsection{The eccentricities and the resonant angles}
Since $J_{\rm sum}$ and $J_{\rm dif}$ are integrals, $J_1=
\frac{1}{2}(J_{\rm sum}+J_{\rm dif})$ and $J_2= \frac{1}{2}(J_{\rm
sum}-J_{\rm dif})$ are invariables too. They constrain the
variation of the planetary semi-major axes by $3J_1+J_2 =
3m_1^\prime\sqrt{\mu_1a_1} +m_2^\prime\sqrt{\mu_2a_2} = const.$
However, when we discuss planets trapped in an MMR, we may simply
assume the semi-major axes are constants, otherwise the resonance
would be destroyed. In fact, the numerical simulations in section
2 also support this assumption (Fig.\,2). Similarly, simple
calculations show that
 \be J_{\rm sum}= m_1^\prime\sqrt{\mu_1a_1(1-e_1^2)}+
 m_2^\prime\sqrt{\mu_2a_2(1-e_2^2)}. \ee
This confines the variation of the eccentricities when we fix the
semi-major axes $a_1$ and $a_2$. Besides this, the variation range
of the eccentricities is also bounded by other constrains, e.g.,
the total energy of the system.

\begin{figure}
\vspace{6cm} \includegraphics{fig4.eps} \caption{The possible variation of
eccentricities of the planets. The dotted curve corresponds to
$J_{\rm sum}= 3.951733\times 10^{-4}$. Results from the numerical
simulations of Fig.\,1a (plus), b (black dot) and c (open circle)
are plotted to compare. }
%\label{fig4}
\end{figure}

We assume the two semi-major axes are fixed, $a_1= 0.115$\,au and
$a_2=0.241$\,au. Then Eq.\,(4) defines a family of curves on the
$(e_1, e_2)$-plane (Fig.\,4). Each curve in Fig.\,4 corresponds to
a definite value of the integral $J_{\rm sum}$. The orbital
elements in Table 1 give $J_{\rm sum}= 3.591733\times 10^{-4}$, so
that the eccentricities $e_1, e_2$ of orbits beginning from these
initial conditions are restricted on the corresponding curve
during the dynamical evolution.

After assuming $a_{1,2}$ fixed, for certain $e_{1,2}$ values, we
can calculate the energy level curves of the Hamiltonian on the
$(\sigma_1, \sigma_2)$-plane. Fig.\,5 shows such contours. At the
first glance, we note the symmetry over both $\sigma_1$ and
$\sigma_2$. This has been observed in our numerical simulations
(recalling $\theta_i = -2\sigma_i$).

\begin{figure}
\vspace{7.0cm} \includegraphics{fig5.eps} \caption{The energy level curves of
the Hamiltonian on the $(\sigma_1, \sigma_2)$-plane. A gray level
code is used to measure energy values, the darker regions indicate
smaller energy value, and vice versa. {\it a}, {\it b}, {\it c},
and {\it d} are the case for $e_1= 0.03, 0.10, 0.15$ and $0.20$
respectively, with $e_2$ obtained from Eq.\,(4). }
%\label{fig5}
\end{figure}

Since the total energy $H$ is an integral, a system starting from
a point on any curve in Fig.\,5 can never jump to other curve with
a different energy. On the other hand, the variation of
eccentricities (Fig.\,4) is constrained by the integral $J_{\rm
sum}$ through Eq.\,(4). Therefore, as the eccentricities vary
following Eq.\,(4), an energy curve on the $(\sigma_1,
\sigma_2)$-plane will extend to an energy surface (e.g. a
cylinder) and the system will evolve on this surface. To
illustrate this, we fix $a_1= 0.115$\,au and $a_2=0.241$\,au, set
$e_1= 0.03$ (and sequently $0.10, 0.15, 0.20$), evaluate the
corresponding $e_2$ from Eq.\,(4) with $J_{\rm sum}=
3.591733\times 10^{-4}$, then calculate contours for different
pairs of $(e_1, e_2)$. Finally, we get a series of section of the
energy surface in Fig.\,5. In this way, we can follow the
dynamical evolution of a system starting from definite initial
conditions.

For example, the thick curves in Fig.\,5, representing an energy
$H= -3.957679 \times 10^{-3}$, keep the closed oblate shape when
$e_1$ evolves from the initial value $0.03$ to $0.20$. The oblate
profile of the curve implies that the angle $\sigma_1$
($\theta_1$) has a larger reachable range than $\sigma_2$
($\theta_2$)during the temporal evolution of a system with initial
conditions located on this curve. The angle $\sigma_1$ could vary
in a range of $\sim (160\degr, 400\degr)$, while $\sigma_2$ is
bounded between $20\degr$ and $60\degr$. Correspondingly, the
resonant angle $\theta_1$ and $\theta_2$ could vary in ranges with
widths of $480\degr$ (circulation) and $80\degr$ (libration),
respectively. This is just the situation of case {\bf b} in
Fig.\,1. Another example indicated by a dashed curve with
$H=-3.957722\times 10^{-3}$ exhibits the possibility of
circulating of both $\theta_1$ and $\theta_2$.

We denote the possible variation of $\sigma_i$ in Fig.\,5 by
$\Delta\sigma_i$. According to the foregoing analysis,
$\Delta\sigma_2<\Delta\sigma_1$. If $\Delta\sigma_1\la 180\degr$,
both $\theta_1$ and $\theta_2$ would librate ($\Delta\theta_2<
\Delta\theta_1\la 360\degr$) and the system could run as Fig.\,1a.
It could run as Fig.\,1b, if $\Delta\sigma_1 \ga 180\degr$ while
$\sigma_2$ is still bounded in a narrow range, say
$\Delta\sigma_2\la 90\degr$. And if $90\degr \la \Delta\sigma_2
\la 180\degr$, Fig.\,1c could be followed. So the measures of such
areas in the whole $(\sigma_1, \sigma_2)$-plane tell the
probabilities of different types of motion. Calculations show the
proportions of these regions in the whole plane, i.e. the
probabilities of motion type {\bf a}, {\bf b} and {\bf c},
approximatively are $10\%, 16\%$ and $4.5\%$, respectively. They
are coarse estimates because we don't count the stability of the a
possible motion. After considering the stability, these estimates
must decrease in different extents.

In a word, given $a_i, e_i$, the dynamical evolution of a system
is determined by the initial values of $\sigma_i$ ($i=1,2$), which
consequently correspond to different energy $H$.

\subsection{The occurrence of apsidal corotation}
When two planets are in a 3:1 MMR, at least one of the three
resonant angles librates around a definite value. Since $\sigma_2$
generally has a smaller libration amplitude than $\sigma_1$, we
can further simplify the Hamiltonian by safely setting $\sigma_2
\equiv \sigma_2^0$. To determine $\sigma_2^0$, we can calculate
the stable stationary solutions of the Hamiltonian equations
\cite{bea03a}, or we can get an estimate from the position of the
Hamiltonian extremum in Fig.\,5. But in this paper, we simply
adopt $\sigma_2^0=320\degr$ ($\theta_2= 80\degr$) from the
numerical results. Recalling the constant semimajor axes and the
constrain on eccentricities, the Hamiltonian can be explicitly
expressed as $H= H(a_2^0, a_1^0, e_2, e_1(J_{\rm sum}, e_2),
\sigma_2^0, \Delta\varpi )$, with only two ``free'' variables
$e_2$ and $\Delta\varpi$. We calculate the contour of such
Hamiltonian on the $(e_2, \Delta\varpi)$-plane and show it in
Fig.\,6. The contour of the velocity of $\Delta\varpi$ is also
calculated and shown in Fig.\,7.

\begin{figure}
\vspace{5cm} \includegraphics{fig6.eps} \caption{The contour of Hamiltonian
on the $(e_2, \Delta\varpi)$-plane. Some Hamiltonian values are
labelled. Dots and squares represent the case {\bf a} and {\bf b}
in Fig.\,1 from numerical integrations. }
%\label{fig6}
\end{figure}

A system approximately satisfying the assumptions mentioned above
will evolve along one of the curves in Fig.\,6. For the inner two
planets in the 55 Cancri, the initial orbital elements listed in
Table 1 ($e_1^0= 0.03, e_2^0= 0.41$) correspond to the right
boundary of the box in Fig.\,6. Two different fates of a system
starting from these initial conditions are distinguished: one is
characterized by a libration of $\Delta\varpi$ around $\sim
250\degr$ and the other a circulation of $\Delta\varpi$. In the
latter case, the eccentricity $e_2$ can reach a smaller value
(correspondingly $e_1$ a higher value). These agree pretty well
with the directly numerical integrations. Such agreements are
shown in Fig.\,6 by a comparing plots of the numerical results.

Fig.\,7 shows the velocity of $\Delta\varpi$. Through this figure,
we can explain how $\Delta\varpi$ evolves. Along the right
boundary of the box, the absolute value of velocity could be very
large, so that $\Delta\varpi$ increases or decreases quickly. Let
us write the perturbing part of the Hamiltonian as $H_1=
\sum_{j=0}^{j_{\rm max}} \sum_{k=0}^{k_{\rm max}} R_{j,k} e_1^j
e_2^k$, where $R_{j,k}$ contains the coefficients $R_{n,j,k,u,l}$
and sums over $n,u,l$ in Eq.(3). By careful calculations, we
obtain
 \ben
 \frac{d\Delta\varpi}{dt}= \frac {\sqrt{1-e_1^2}}{e_1^2L_1}
 \sum_{j=1}^{j_{\rm max}} \sum_{k=0}^{k_{\rm max}}jR_{j,k}e_1^j e_2^k
 \nonumber \\ -\frac{1-e_2^2}{e_2^2L_2}\sum_{j=0}^{j_{\rm max}}
 \sum_{k=1}^{k_{\rm max}}kR_{j,k} e_1^j e_2^k.
 \een
Obviously, when $e_i \ll 1$, the dominating term on the right hand
side has the order $\sim \frac{1}{e_i}$. Near the right boundary
of the box in Fig.\,7, we have $e_1 \sim 0.03 \ll 1$. That's the
reason why $\Delta\varpi$ has a quick change when $e_1$ approaches
its minimum value as shown in Fig.\,1b,c.

The numerical simulations reveal that the quick varying of
$\Delta\varpi$ in Fig.\,1b,c is caused by a quick varying of
$\varpi_1$ when $e_1\rightarrow 0$. In fact, when an orbit is
nearly circular ($e\sim 0$), it would be relatively easy to change
its direction ($\varpi$), thus the two planets could loose the
locking of periastrons if they were not in a strong coupling. On
the other hand, whether they can strongly couple with each other
depends also on how close they can approach each other, that is,
how large their eccentricities (especially $e_2$) are. Bearing
this in mind we can understand why in case {\bf a} $e_2$ has a
higher lower-limit than in cases {\bf b} and {\bf c}. Of course, a
small $e_2$ can also cause a quick variation of $\Delta\varpi$,
but this situation does not happen because the energy integral
prevents $e_2$ from approaching very small value.

\begin{figure}
\vspace{5cm} \includegraphics{fig7.eps} \caption{The contour of ${d
\Delta\varpi}/{dt}$. Some velocity values are labelled and the
scale is $360\degr\,{\rm yr}^{-1}$. Dashed curves indicate the
zero velocity curve. Dots and squares are the same as in Fig.\,6.}
%\label{fig7}
\end{figure}

From Fig.\,6 and Fig.\,7, we see different behaviors of
$\Delta\varpi$ (circulation or libration) have different varying
directions along the right edge of the box. By numerically solving
the equation $d\Delta\varpi/ dt=0$, we find two zero velocity
points on the boundary. They define the boundaries of the initial
conditions leading to a librating or circulating of
$\Delta\varpi$. The calculations show that the apsidal corotation
happens if $\Delta\varpi^0\in (160\degr, 330\degr)$, while if
$\Delta\varpi^0\in (-30\degr, 160\degr)$, $\Delta\varpi$
circulates.

The case {\bf c} has a Hamiltonian value a little further from the
extremum in Fig.\,5, so that $\sigma_2 (\theta_2)$ has a large
amplitude of libration and can not be assumed as a constant. As a
result, we can not find a corresponding curve for it in Fig.\,6.
However, if we draw a series of Hamiltonian contour with different
value of $\sigma_2$ (adding another dimension to Fig.\,6), we can
also understand its dynamical evolution.

Fig.\,8 summarizes the initial conditions of the 38 stable
examples in the numerical simulations. For clarity, we have mapped
the symmetric configurations to one regime according to the
symmetry of the Hamiltonian. Except the three examples as case
{\bf c}, all the points for stable systems gather along the line
of $\sigma_2^0=320\degr$, and points for librating and circulating
$\Delta\varpi$ (case {\bf a} and {\bf b}) are separated by the
dotted lines of $\Delta\varpi^0=160\degr, 330\degr$. These agree
very well with the above analysis with the Hamiltonian.

\begin{figure}
\vspace{5cm} \includegraphics{fig8.eps} \caption{The initial conditions of
the 38 stable systems. Open squares, plus and solid squares
indicate the systems evolving as case {\bf a}, {\bf b} and {\bf c}
in Fig.\,1. The dotted lines give the positions of
$\Delta\varpi^0= 160\degr, 330\degr$ and the solid line is
$\sigma_2^0=\Delta\varpi^0 +\sigma_1^0 = 320\degr$.}
%\label{fig8}
\end{figure}

We have given an estimate of the probability of following
different ways as case {\bf a, b} and {\bf c}. The analyses here
tell that all the initial conditions leading to stable
configurations gather in a narrow strip along the line in Fig.\,8.
That is, for systems having given $a_{1,2}, e_{1,2}$ as in Table
1, only those with initial conditions satisfying
$\sigma_2^0\approx 320\degr$ and $\sigma_1^0+\Delta\varpi^0\approx
320\degr$ can survive for long time.

\subsection{Surfaces of section}
We know now the initial conditions determine whether the apsidal
corotation happens or not, but whether a system would follow a
definite evolution type also depends on the stability of an
evolution type. The stability of an orbit in the phase space can
be revealed by its projection on the surface of section. We
present in Fig.\,9 the sections of $(I_2\cos\sigma_2,
I_2\sin\sigma_2)$ for the three orbits in Fig.\,1. The plane is
defined as $\sigma_1=270\degr, \dot{\sigma}_1>0$. To obtain these,
we adopt the initial conditions of orbits from Fig.\,1, and then
integrate the corresponding Hamiltonian equations based on the
expansion as Eq.\,(2,3).

\begin{figure*}
\vspace{6.0cm} \includegraphics{fig9.eps} \caption{The projection on the
surfaces of sections for case {\bf a}, {\bf b} and {\bf c} in
Fig.\,1.}
%\label{fig9}
\end{figure*}

Fig.\,9c indicates case {\bf c} has a chaotic orbit. On the
contrary, the invariant curves for cases {\bf a} and {\bf b}
(Fig.\,9a,b) imply that both these two orbits are regular. That
is, no matter whether the apsidal corotation happens or not, the
system trapped in the 3:1 MMR could be stable. From this point of
view we may argue the apsidal corotation only has a limited
contribution to the stability of the system.

Because the Hamiltonian is expressed explicitly in $a_i,e_i$ and
the integrals of motion $J_{\rm sum}$, $J_{\rm dif}$ affect the
system as constrains, we can only define an energy surface ($H
\equiv H_0$) with certain pre-determined $J_{\rm sum}, J_{\rm
dif}$, then calculate the surface of section. We find the
structure of the phase space and the stability of the system
depends sensitively not only on $H$ but also on $J_{\rm sum}$ and
$J_{\rm dif}$. For example, with the same values of $H, J_{\rm
sum}$ and $J_{\rm dif}$ as the ones in Fig.\,1b, we calculate tens
of orbits with different initial conditions. On the surface of
section, these orbits occupy a big regular-motion region. This
result does not conflict with the fact that there are only a small
fraction of stable systems in numerical simulations, in fact, when
we change $H$, (or $J_{\rm sum }$, $J_{\rm dif}$) just a little,
we find remarkable chaotic region on the surface of section. The
application of the surface-of-section technique is limited by the
high dimension of the system. This will be discussed in detail in
our future paper.

\section{Summary and Discussion}
With hundreds of numerical simulations of the planetary system of
the 55 Cancri, we find the third planet has a very weak influence
on the motion of the inner two planets. We confirm the inner two
could be trapped in a 3:1 mean motion resonance and three
different types of motion are found. Judging from the Lyapunov
character indicators and the surviving time of integration, two of
them (case {\bf a, b}) are practically stable, so that the real
system could be running in one of these configurations.

Via a new Hamiltonian expansion which is suitable for
high-eccentricities planar three-body problem, we study the
dynamics of the different configurations. We discuss the
variations of eccentricities and resonant angles, explain the
happenings of different evolving types, and give a criterion of
the occurrence of the apsidal corotation.

The surfaces of section for the three types of motion are
calculated and they reveal the stabilities of systems with or
without the apsidal corotation. With these results we argue that
the stability of the system is mainly due to the 3:1 MMR, and the
apsidal corotation has only a limited contribution. We would like
to mention that this method can also be applied to other
extra-solar planetary systems with other mean motion resonances.

Numerical simulations suggest at least $\sim10$ percent of systems
are in a 3:1 MMR, while the Hamiltonian analyses give an upper
limit of $\sim 30\%$, which however will drop down after
considering the stability. We also list the initial conditions
leading to this MMR.

Systems with different motion configurations have different energy
($H$) levels. The apsidal corotation happens when the Hamiltonian
approaches the extreme value. So, if the two planets are captured
into current configuration through orbital migration caused by the
action of non-conservative forces, the system should have an
extremum of energy so that the apsidal corotation happens
\cite{kle03}. When the future observations would reveal more
accurate properties of this system, we may consider what
signatures of the migration are still presented in this system.

The masses of planets adopted in this paper are the values from
the orbital solutions when assuming $\sin i=1$. As for the
situations of $\sin i<1$, our initial analysis with the
Hamiltonian get the very similar results in a wide range of $\sin
i$. This is also consistent with the results in \cite{bea03a}. We
have also analysed the possible motion configurations and their
stabilities if the initial eccentricities differ from the values
adopted above. With the help of the Hamiltonian model, we find
that a higher $e_2$ favors the establishment of a 3:1 MMR. Anyway,
we'd like to leave more details of these to our future paper.

Last but not least, the general relativity effect may affects the
secular dynamics of the 55 Cancri system, since the Companion B
and C are quite close to the central star. For the innermost
planet, the orbital precession caused by the general relativity
effect, is calculated. Although this periastron shift,
$\sim1.66\times 10^{-6}$ radians per orbit, is quite small, it's
about three times larger than that of Mercury in our Solar system.

\section*{Acknowledgements}
We thank Dr. Beaug\'e for helpful discussions. Appreciations also
go to Dr. Tsevi Mazeh for the constructive comments and
suggestions. This work is supported by Academy of Finland (Grant
No.\,44011). LYZ and YSS are also supported by the Natural Science
Foundation of China (No.\,10233020) and the Special Funds for
Major State Basic Research Project (G200077303).

\end{document}